\documentclass[showpacs, pra,twocolumn,preprintnumbers ,amsmath, amssymb, superscriptaddress, aps]{revtex4-2}
\usepackage{amsfonts}
\usepackage{color}
\usepackage{amsmath,amssymb}
\usepackage{pifont}
\usepackage{amssymb}  
\usepackage{bbold}
\usepackage{float}
\usepackage{subfloat}

\usepackage[caption=false]{subfig}
\usepackage{tikz}
\usepackage{makecell}
\usepackage{subfig}
\usepackage{pifont}   
\usepackage{graphicx} 
\graphicspath{{Figures/}}
\usepackage{dcolumn}  
\usepackage{bm}       
\usepackage{multirow} 
\usepackage{placeins}
\usepackage[colorlinks]{hyperref}
\usepackage{mathtools}
\usepackage{appendix}

\def \be{\begin{align}}
	\def \ee{\end{align}}
\def \bea{\begin{eqnarray}}
	\def \eea{\end{eqnarray}}

\begin{document}

	\title{
		{Aharonov-Bohm flux and dual gaps effects on  energy levels in graphene magnetic quantum dots}}
	\date{\today}
	\author{Fatima Belokda}
	\affiliation{Laboratory of Theoretical Physics, Faculty of Sciences, Choua\"ib Doukkali University, PO Box 20, 24000 El Jadida, Morocco}
	\author{Ahmed Bouhlal}
	\affiliation{Laboratory of Theoretical Physics, Faculty of Sciences, Choua\"ib Doukkali University, PO Box 20, 24000 El Jadida, Morocco}
	\author{Ahmed Siari}
	\affiliation{Laboratory of Theoretical Physics, Faculty of Sciences, Choua\"ib Doukkali University, PO Box 20, 24000 El Jadida, Morocco}
	\author{Ahmed Jellal}
	\email{a.jellal@ucd.ac.ma}
	\affiliation{Laboratory of Theoretical Physics, Faculty of Sciences, Choua\"ib Doukkali University, PO Box 20, 24000 El Jadida, Morocco}
	\affiliation{Canadian Quantum Research Center, 204-3002 32 Ave Vernon,  BC V1T 2L7, Canada}


\begin{abstract}	
	
	We address the question of how the Aharonov-Bohm  flux $\Phi_{AB}$ can affect the energy levels of graphene magnetic quantum dots (GMQDs) of radius $R$. To answer this question, we consider GMQDs induced by a magnetic field $B$ and subjected to two different gaps - an internal gap $\Delta_1$ and an external gap $\Delta_2$. After determining the eigenspinors and ensuring continuity at the boundary of the GMQDs, we formulate an analytical equation describing the corresponding energy levels. Our results show that the energy levels can exhibit either a symmetric or an asymmetric behavior depending on the valleys $K$ and $K'$ together with the quantum angular momentum $m$. In addition, we find that $\Phi_{AB}$ causes an increase in the band gap width when $\Delta_1$ is present inside the GMQDs. This effect is less significant when a gap is present outside, resulting in a longer lifetime of the confined electronic states. Further increases in \(\Phi_{AB}\) reduce the number of levels between the conduction and valence bands, thereby increasing the band gap. These results demonstrate that the electronic properties of graphene can be tuned by the presence of the AB flux, offering the potential to control the behavior of graphene-based quantum devices.
            
	\vspace{0.25cm}
	
	
\end{abstract}
\maketitle
	\section{Introduction}
\par 
It is the hexagonal arrangement that gives graphene its unique properties \cite{Novs}. These include exceptional electronic conductivity, record thermal conductivity, high mechanical strength, flexibility, and transparency \cite{geim2007rise, xia2010index, guinea}, making it one of the most robust materials \cite{kats,xia}. Graphene has a peculiar electronic band structure and is characterized by the presence of singular points called Dirac points, which are at the Fermi level. This means that there is no direct gap between the valence and conduction bands \cite{hwang2008acoustic, neto}. The Dirac equation provides valuable insights into the properties of relativistic massless particles that underlie the linear behavior observed in graphene \cite{sprinkle,zhan}.

The need for a band gap has led to the development of several techniques to remotely control the conductance in graphene. Among these techniques is the confinement of fermions in graphene to create quantum dots (QDs). Numerous methods for achieving this confinement have been documented in the literature, one see the review paper \cite{rozh}. These include the application of inhomogeneous magnetic fields \cite{martin,espino}, cylindrical potential symmetry \cite{chen}, and spatial modulation of the Dirac band gap \cite{giav}. The confinement can also be realized by carving the flake into small nanostructures \cite{zebro,thom}, inducing a band gap by substrate interaction \cite{recher,pablo}, and using a co-precipitation method \cite{saeed,kumar}.
We recall that QDs offer innovative applications in electronic devices \cite{berger}, valves \cite{espi}, photovoltaics \cite{bacon}, qubits \cite{trauz}, and gas detection \cite{sun}, among others. The electronic properties of graphene quantum dots has been studied in several occasions.

We are interested in studying how the Aharonov-Bohm (AB) flux $\Phi_{AB}$ affects the energy levels of graphene magnetic quantum dots (GMQDs) of radius $R$ and thus modifies the band gap. More specifically, we want to understand how $\Phi_{AB}$ together with both the external band gap $\Delta_2$ and the internal band gap $\Delta_1$ affect the electronic bands. By solving the Dirac equation, we obtain the eigenspinors as a function of $\Phi_{AB}$, $\Delta_1$, and $\Delta_2$ for the valleys $\tau = \pm1$. Then we determine the corresponding energy levels by ensuring the continuity of the eigenspinors at the boundary of the quantum dots.
Among the results, we observe both symmetric and asymmetric behavior in the energy levels. In particular, the flux changes the electronic band structure of the quantum dots. Specifically, we show that the incorporation of $\Phi_{AB}$ widens the band gap, creating a distinction between the valence and conduction bands while maintaining the symmetry of the energy levels. This broadening extends the trapping time of electrons within the GMQDs. Furthermore, the introduction of an energy band further widens the gap and introduces new energy levels, albeit breaking the energy symmetry. In summary, we argue that the AB flux serves as an external source to control the electronic properties of our system.

The paper is organized as follows: In Sec.~\ref{MMM}, we present the theoretical framework relevant to the present system and solve the eigenvalue equation to obtain the eigenspinors. Matching these at the boundary of the GMADs yields the formula for the energy levels as a function of $B$, $R$, $\Phi_{AB}$, $\Delta_1$, and $\Delta_2$. In Sec.~\ref{DDD}, due to various dependencies, we perform a numerical analysis of the energy levels under appropriate conditions. Finally, we conclude our results. 

	\section{Theoretical Model} \label{MMM}
	
\begin{figure}[H]
		\centering
		\includegraphics[scale=0.5]{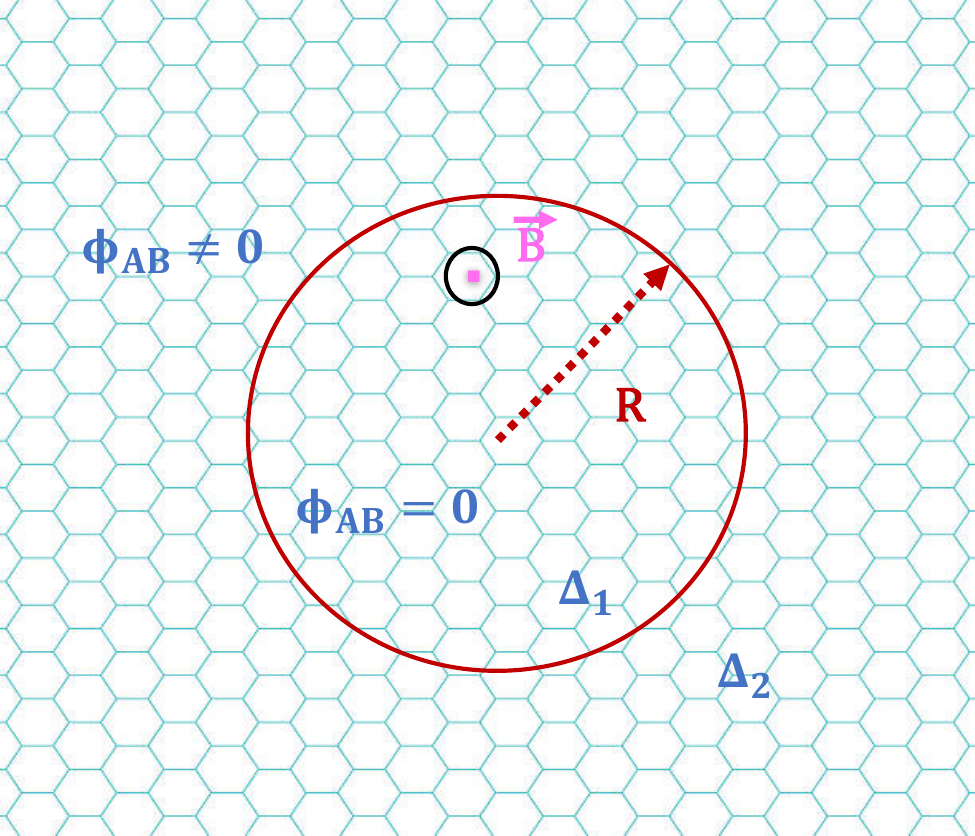}
		\caption{(color online) Schematic profile of a graphene quantum dot with radius \( R \) generated by a magnetic field \( B \) that is influenced by Aharonov-Bohm flux fields \(\Phi_{\text{AB}}\) and dual gaps \(\Delta_i, i=1,2\).}\label{system}
	\end{figure}

	We consider a system of graphene magnetic quantum dots (GMQDs) subjected to the Aharonov-Bohm (AB) flux and dual gaps, as shown in Fig. \ref{system}. The introduction of these two quantities can have significant effects on the electronic properties of MGQDs. The AB effect is a quantum mechanical phenomenon that occurs when charged particles are affected by a magnetic field, even in regions where the magnetic field strength is zero. This effect occurs due to the phase acquired by the wave function when charged particles traverse a closed path in the presence of a magnetic flux. In our case, the AB effect is produced by the magnetic flux passing outside the MGQD \cite{Aharonov1959}. The combination of the AB flux and dual gaps introduces additional complexity into the system, resulting in unique electronic properties. The AB effect can modify the energy spectrum, wave functions, and transport properties of the fermions in the graphene quantum dot. The dual gaps also contribute to the spatial variation of the energy levels and can lead to localized states or resonances within the dot \cite{Myoung19, Belokda}.  
	The present study contributes to our understanding of novel electronic phenomena and their potential applications in areas such as in bioimaging, biosensing, and therapy \cite{Chung2021}.
	
The magnetic field $\vec{B} = B(r) \vec{e}_z$, which ensures the confinement of the fermions, is chosen to be perpendicular to the $(x,y)$-plane and is given by
	\begin{equation}
		B(r) = \begin{cases}B, & r \leq R \\ 0, & r>R\end{cases}
	\end{equation}
where $\vec{e}_z$ is the unit vector. Actually, the resulting vector potential is composed of two components: $\vec{A} = \vec{A_1} + \vec{A_2}$ \cite{ikhdair2015nonrelativistic}, where $\vec{A_1}=G(r) \vec{e_\theta}$ and $\vec{A_2}=\frac{\phi_{AB}}{2 \pi r} \vec{e_\theta}$ is related to the applied AB flux, such as
	\begin{equation}
		G(r)= \begin{cases}\frac{B}{2 r}\left(r^2-R^2\right), & r\leq R \\ 0, & r>R \end{cases}, \quad  \phi_i=\frac{\phi_{AB}}{\phi_0}
	\end{equation}
and $\phi_0 = h/e$ is unit flux. We study the energy spectrum of Dirac fermions in the honeycomb lattice of covalently bonded carbon atoms in graphene, subjected to both a magnetic field and an AB flux, by considering the Hamiltonian
	\begin{equation}\label{hamilt}
		H_\tau=v_F\left(\pi_x \sigma_x+ \tau \pi_y \sigma_y\right) +\tau \Delta_i(r) \sigma_z
	\end{equation}
	where $v_F=10^6 \mathrm{~m} / \mathrm{s}$ denotes the Fermi velocity, $\tau= \pm $ denotes the valleys $K$ and $K^{\prime}$, $\pi_i=p_i+e A_i$ represents the kinetic momentum operator, and $\sigma_i (i=x, y, z)$ are the Pauli matrices in the basis of the two sublattices of $A$ and $B$ atoms. The non-uniform gap $\Delta_i(r)$ is given by
	\begin{equation}
		\Delta_i (r) = \begin{cases}
			\Delta_1 = \hbar v_F \delta_1, & r<R\\ \Delta_2 = \hbar v_F \delta_2, &   r>R.
		\end{cases}
	\end{equation}
The system under consideration has a spherical geometry (see Fig.\ref{system}), then we introduce the polar coordinates $(r, \theta)$. Consequently, we can express \eqref{hamilt} as
		\begin{equation}
		H^{\tau}=\hbar v_F\begin{pmatrix}
			\tau \delta_i & \pi^{+} \\
		\pi^{-}	 & -\tau \delta_i
		\end{pmatrix}
		\end{equation}
where we have define the operators 
		\begin{equation}
	\pi^{\pm}	=	 e^{\mp i \tau \theta}\left[-i {\partial_ r}\mp i  \tau \left(\frac{i}{r} {\partial_ \theta}+ \frac{e B r}{2}+\frac{e\phi_{AB}}{2\pi r}\right)\right].
		\end{equation}

As long the total momentum operator $J_z=L_z+\hbar \sigma_z / 2$ commute with the Hamiltonian \eqref{hamilt}, the eigenstates $	\Psi^{\tau}(r, \theta)$ ca be represented by
\begin{equation}
			\Psi^{\tau}(r, \theta)=e^{i m \theta}\binom
					{\varphi^{\tau}_A(r)} 
					{i e^{i \tau \theta} \varphi^{\tau}_B(r)}
	\end{equation}
where $m\in\mathbb{Z}$ is a quantum number related to $J_z$. Using the eigenvalues $H^{\tau} \Psi^{\tau}(r, \theta) = E \Psi^{\tau}(r, \theta)$, we derive two coupled equations within the region $r<R$ (inside the GMQDs)
		\begin{align} \label{eq8}
				&  \left[{\partial_ \rho} + \frac{\tau}{\rho}\left(m  + \tau \right) +\tau \beta\rho \right] \varphi^{\tau}_{B}(\rho)=(\epsilon-\tau \delta_1) \varphi^{\tau}_{A}(\rho) \\\label{eq9}
				&  \left[ - {\partial_\rho} + \frac{\tau m}{\rho}  +\tau \beta\rho \right] \varphi^{\tau}_{A}(\rho)=(\epsilon+\tau \delta_1) \varphi^{\tau}_{B}(\rho)
			\end{align}	
 where  we have introduced  dimensionless parameter dimensinless parameters $\rho=\frac{r}{R}$,  $\delta_{i}={\frac{\Delta_{i}R}{\hbar\nu_{F}}}$, and $\epsilon={\frac{E}{\hbar\nu_{F}}}$, with $ \beta=\frac{eBR^{2}}{2\hbar}$.
 By injecting  \eqref{eq9} into  \eqref{eq8} to find a second  order differential equation for $\varphi^{\tau}_{A}(r)$
 \begin{equation} \label{eqdif}
 			\left[{\partial^2_\rho} + \frac{1}{\rho} {\partial_ \rho}-\frac{m^2}{\rho^2} - \rho^2{\beta} - {2(m + \tau)}{\beta}+k_1^2\right] \varphi^{\tau}_{A}(\rho) =0
 		\end{equation}
and $k_{1}=\sqrt{|\varepsilon^{2}-\delta_{1}^{2}|} $ is the wave vector.
To solve \eqref{eqdif}, we first examine the asymptotic limits that describe the required physical behavior depending on the value of $\rho$. Since $\rho \rightarrow \infty$, we can approximate \eqref{eqdif}  as
\begin{equation}\label{e:10}
	\left( \partial_\rho^2+\frac{1}{\rho}\partial_\rho - \rho^2 \beta\right)  \varphi^{\tau}_{A}(\rho)=0
\end{equation}
which is of zero-order Bessel type. Thus, the solution is expressed as
\begin{equation}
	\varphi^{\tau}_{A}(\rho) =c_1 I_0\left( \rho^2 \beta\right) +c_2 K_0\left( \rho^2 \beta\right) 
\end{equation}
where $I_0(x) $ and $K_0(x) $ represent the modified zero-order Bessel functions of the first and second types, respectively. We choose $c_1 = 0$ and $c_2 = 1$ to prevent the function $I_0(x) $ from diverging as $x $ approaches infinity. Using the asymptotic behavior of $K_0(x) \underset{x\gg 1}{\sim} \frac{e^{-x}}{\sqrt{x}}$, we approximate $\Phi^A(r)$ as
\begin{equation}\label{e:13}
	\varphi^{\tau}_{A}(\rho) \sim e^{\frac{-\beta \rho^2}{2}}.
\end{equation}
In the limit $\rho \rightarrow 0$,  \eqref{eqdif} reduces to	
\begin{equation}\label{e:14}
	\left(  \partial_\rho^2+\frac{1}{\rho}\partial_\rho  - \frac{m^2}{\rho^2} \right) \varphi^{\tau}_{A}(\rho)=0
\end{equation}	
which has the following  solution	
\begin{equation}\label{e:15}
	\varphi^{\tau}_{A}(\rho) =\frac{c_3}{2}(\rho^m +\rho^{-m})+\frac{i c_4}{2}(\rho^m -\rho^{-m})
\end{equation}	
To ensure that the solution satisfies the physical constraints, $c_3$ and $c_4$ must be chosen appropriately. Therefore, we consider the cases of positive and negative values of $m$ separately. For $m\ge 0$, where $\sim \rho^{-m}$ must vanish, we set $c_4 = -ic_3$, and by convention we assign $c_3 = 1$. Conversely, for $m< 0$, where $\sim \rho^m$ must vanish, we replace $c_4 = ic_3$ with $c_3 = 1$. Combining these considerations, we express the asymptotic behavior of $\varphi^{\tau}_{A}(\rho)$ as
\begin{subequations}\label{16}
	\begin{align}
		&\varphi^{\tau}_{A}(\rho) \sim \rho^m ,\qquad m\ge 0, \label{16a}\\
		&\varphi^{\tau}_{A}(\rho) \sim \rho^{-m},\qquad m<0 \label{16b} .
	\end{align}
\end{subequations}
Now it is clear that  the solution of \eqref{eqdif} can be written as 
	 \begin{equation} \varphi_{A}(\rho)=\rho^{|m|}e^{-\frac{\beta\rho^{2}}{2}}\xi(\beta\rho^{2})\label{eqanst}
 \end{equation} and define a new variable $\zeta=\beta\rho^{2}$. Substituting \eqref{eqanst} into \eqref{eqdif}, we obtain the confluent hepergeometric ordinary differential equation
 \begin{equation} \left[\zeta\partial^{2}_\zeta+(p-\zeta)\partial_\zeta-q\right]\xi(\zeta)=0 
 \end{equation}where we have defined
 \begin{equation}   
 	p=1+|m|, \quad            q=-\frac{k^{2}_1}{4\beta}+\frac{|m|+m+\tau+1}{2} \end{equation}
  Consequently the solution are the confluent hypergeometric functions \cite{Abra}. We combine all to get the eigenspinors inside the GMQDs
  \begin{widetext}
  \begin{equation}   \psi^{\tau}_{1}(\rho,\theta)= C^{\tau}_{1}\rho^{|m|}e^{\frac{\beta\rho^{2}}{2}}e^{im\theta}\begin{pmatrix} U(q,p,\beta\rho^{2})  \\\frac{ie^{i\tau\theta}}{\varepsilon+\tau\delta_{1}}\left[\frac{-|m|+\tau m}{\rho}+(\tau+1)\beta\rho\ U(q,p,\beta\rho^{2})-2\beta\rho\frac{q}{p}\ U(q+1,p+1,\beta\rho^{2})\right]\end{pmatrix}.
  \end{equation}	
  \end{widetext}

   In region  $r>R$ (outside the GMQDs),  \eqref{eqdif} can be represented by
  \begin{equation} \label{eqdif2}
 	\left[{\rho^{2}\partial \rho^2} + \rho\partial \rho-(m+\phi_i)^2- k_2^2 \rho^{2}\right] \varphi^{\tau}_{A}(\rho) =0
 \end{equation}
 with the wave vector $k_{2}=\sqrt{|\varepsilon^{2}-\delta_{2}^{2}|} $.
This differential equation has  the Bessel function of the first kind as solution
 \begin{equation}
 	\varphi^{\tau}_{A}(\rho)=C^{\tau}_{2}J_{m+\phi_i}(k_{2}\rho)
 	\end{equation} 
 and $C^{\tau}_{2}$ is the normalization constant. Now, we can use \eqref{eq8} to derive  the second component. By combing all, we  write the eigenspinors as
 \begin{widetext}
 	\begin{align}
 	\psi^{\tau}_{2}{(\rho,\theta)}= C^{\tau}_{2}e^{im\theta} 
 	\begin{pmatrix} J_{m+\phi_i}(k_{2}\rho)  \\\frac{ie^{i\tau\theta}}{\varepsilon+\tau\delta_{2}}[-k_{2}\ J_{m+\phi_i-1}(k_{2}\rho)+\frac{(m+\phi_i)(\tau+1)}{\rho}\ J_{m+\phi_i}(k_{2}\rho)] \label{22}
 	\end{pmatrix}.
 \end{align}
  \end{widetext}
 
Since we are unable to obtain the eigenvalues explicitly, we proceed by formulating an equation that dictates their behavior. Specifically, we use the continuity of both solutions (inside and outside the point) at the boundary $ r= R$ or $\rho=1$, denoted as $\psi^{\tau}_{1}(1)=\psi^{\tau}_{2}(1)$. This approach results in the following equation:
\begin{widetext}
\begin{align}\begin{split}  \frac{J_{m+\phi_i}(k_{2})}{\varepsilon+\tau\delta_{1}}\left[-|m|+\tau m+(\tau+1)\beta\ U(q,p,\beta)-2\beta\frac{q}{p}\ U(q+1,p+1,\beta)\right]-\\
		\frac{U(q,p,\beta)}{\varepsilon+\tau\delta_{2}}\left[-k_{2}\ J_{m+\phi_i-1}(k_{2})+(m+\phi_i)(\tau+1)\ J_{m+\phi_i}(k_{2})\right]=0 \label{eq19}. \end{split}  
\end{align}	
\end{widetext}
 The analytical solution of equation \eqref{eq19} is complex and requires the use of a numerical approach. We will plot the dependence of the energy spectrum on the physical parameters characterizing the GMQDs. This will help us to explore the fundamental properties of the GMQDs and underline their basic characteristics.

 	\section{Discussions}\label{DDD}
 		
\begin{figure*}[ht]
 			\centering
 			\includegraphics[width=5.5cm]{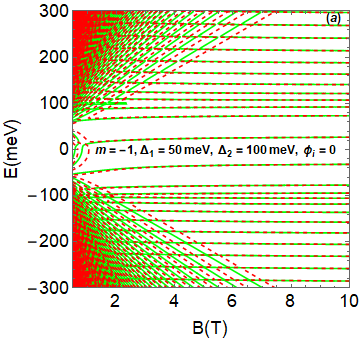}	
 			\includegraphics[width=5.5cm]{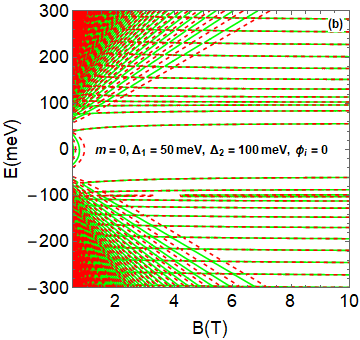}
 			\includegraphics[width=5.5 cm]{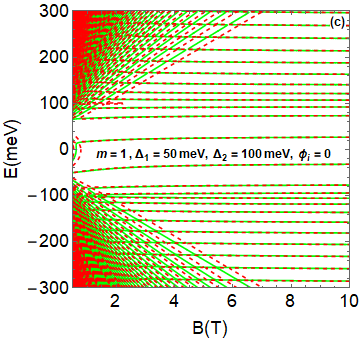}\\
 			\includegraphics[width=5.5cm]{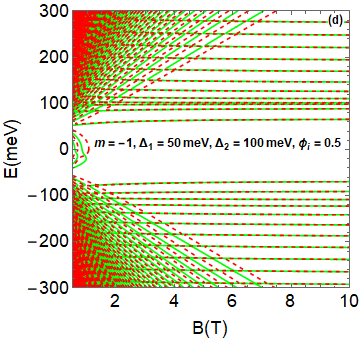}
 			\includegraphics[width=5.5cm]{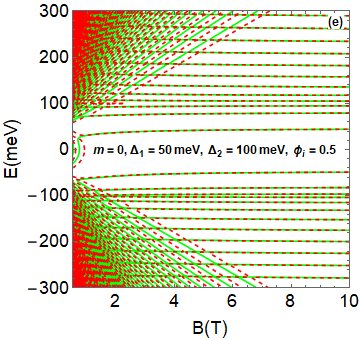}
 			\includegraphics[width=5.5cm]{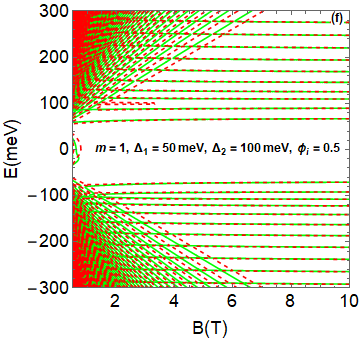}\\
 			\includegraphics[width=5.5cm]{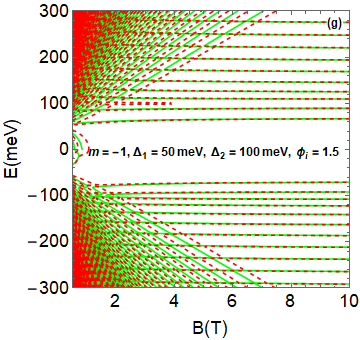}
 			\includegraphics[width=5.5cm]{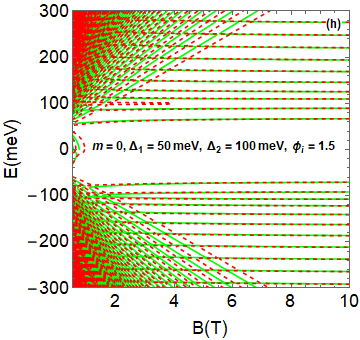}
 			\includegraphics[width=5.5cm]{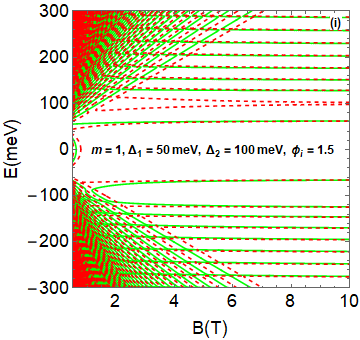}\\
 			\caption  {(color online) Energy levels plotted against the magnetic field $B$ for a radius  $R=70$ nm, dual gaps: $\Delta_{1}=50$ meV and $\Delta_{2}=10$ meV, quantum number $m$: (a,d,g): -1, (b,e,h): 0, (c,f,i): 1,   AB flux $\phi_i$:    (a,b,c):0, (d,e,f):  0.5,  (g,h,i): 1.5, with green curves for $\tau=-1$, and  dotted red curves for $\tau=1$. 
 			}	\label{f2} 
 		\end{figure*}

 Fig. \ref{f2} shows the energy levels versus the magnetic field $B$ for a radius of $R=70$ nm and dual gaps, $\Delta_{1}=50$ meV and $\Delta_{2}=2 \Delta_{1}=100$ meV.  In Figs. \ref{f2}(a,b,c) for zero AB flux $\phi_i=0$, there is emergence of continuous energy bands for three quantum numbers $m=0,\pm 1$ as $B$ approaches zero ($B \rightarrow 0$). As a consequence, the energy states in the valleys $K$ ($\tau=+$) and $K'$ ($\tau=-$) become degenerate. 
 We observe that the increase in $B$ removes the degeneracy and breaks the symmetry $E(m,\tau)\neq E(m,-\tau)$ due to the appearance of an induced band gap. This is consistent with previous results reported in \cite{Belokda, Mirza, Farsi21, Myoung19}.
 This symmetry breaking is analogous to the Zeeman effect, where energy levels are split based on spin orientation relative to the magnetic field. 
At low energies, minor energy levels appear in addition to the main energy bands, and as $B$ grows, linear levels appear. This increases the band gap between the valence and conduction bands. 
 Figs. \ref{f2}(d,e,f) show that the symmetry of energy levels remains broken even in the presence of $\phi_i$. Furthermore, as $\phi_i$ increases, one sees that emergence of an additional band gap. This later causes a further separation between the energy levels occupied by the electrons and those occupied by the corresponding holes. This trend becomes more pronounced in Figs. \ref{f2}(g,h,i) for $\phi_i=1.5$. 
     
 		
\begin{figure*}[ht] 
 	\centering
 	\includegraphics[width=5.5cm]{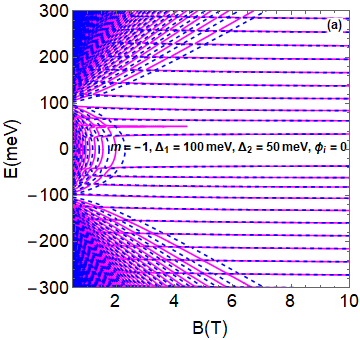}	
 	\includegraphics[width=5.5cm]{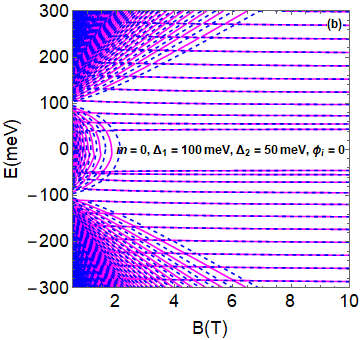}
 	\includegraphics[width=5.5 cm]{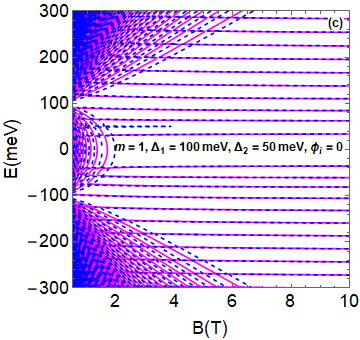}\\
 	\includegraphics[width=5.5cm]{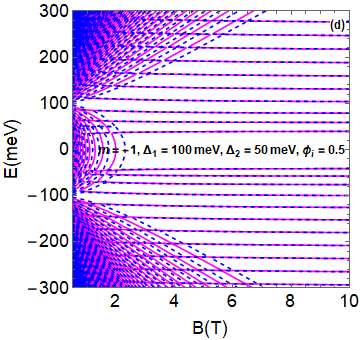}
 	\includegraphics[width=5.5cm]{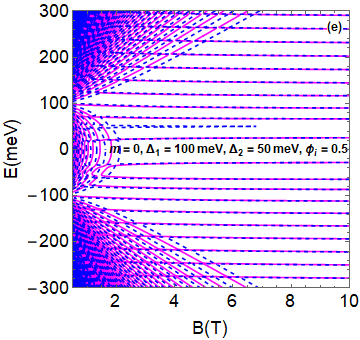}
 	\includegraphics[width=5.5cm]{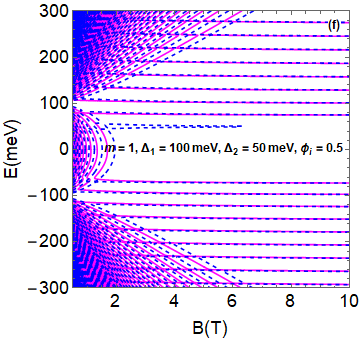}\\
 	\includegraphics[width=5.5cm]{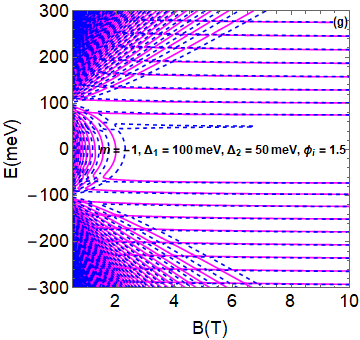}
 	\includegraphics[width=5.5cm]{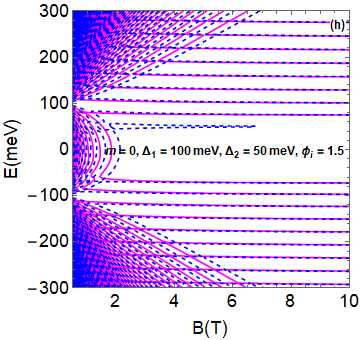}	\includegraphics[width=5.5cm]{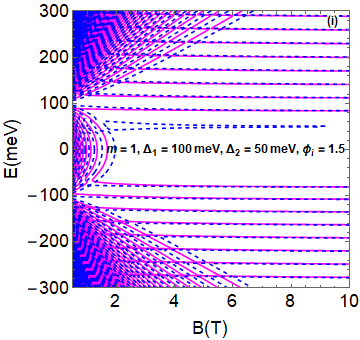}\\
 	\caption  {(color online) The same as in Fig. \ref{f2} but by inverting  dual gaps: $\Delta_{1}=100$ meV and $\Delta_{2}=50$ meV, with magenta curves for $\tau=-1$, and  dashed blue curves for $\tau=1$. 
 	}\label{f3}
 \end{figure*}

 		With the same conditions as in Fig. \ref{f2}, we simply invert the dual gap values $\Delta_{1}=100$ meV and $\Delta_{2}=50$ meV in Fig. \ref{f3}. We observe that the energy levels behave similarly to Fig. \ref{f2}, but there are notable differences. For small $B$, new energy levels with vertical parabolic shapes appear, which are approximately confined in the energy range between $-\Delta_{1}$ and $\Delta_{1}$.
 However, as $B$ increases, these additional levels gradually disappear, the energy degeneracy decreases, and the emerging energy levels break the symmetry. This leads to a significant widening of the band gap between the valence and conduction bands, as shown in Figs. \ref{f3}(a,b,c) for $\phi_i=0$. Now, in the presence of $\phi_i$, we find that the symmetry of the system remains broken, creating an additional band gap, which is clearly visible in Figs. \ref{f3}(d,e,f,g,h,i). 
 Contrary to previous results \cite{Belokda}, the introduction of $\phi_i$ has a significant effect on the quantum confinement in GMQDs. It reduces the presence of inter-band linear levels and increases the width of the band gap. Consequently, the AB flux affects the electronic properties of the fermions in the GMQDs.

\begin{figure*}[ht]
	\centering
	\includegraphics[width=5.5cm]{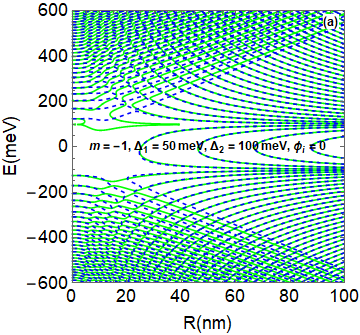}	
	\includegraphics[width=5.5cm]{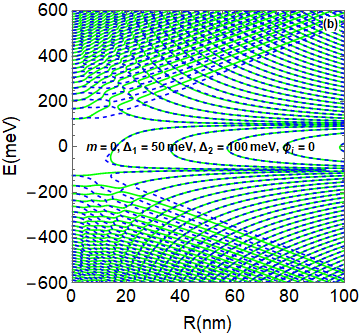}
	\includegraphics[width=5.5 cm]{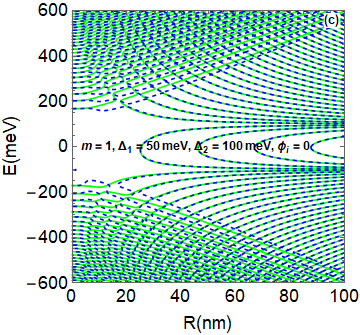}\\ 
	\includegraphics[width=5.5cm]{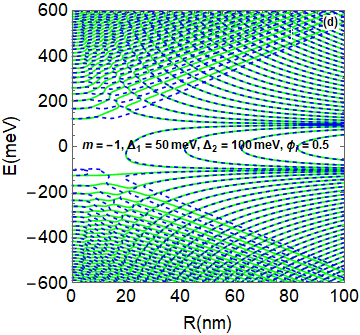}
	\includegraphics[width=5.5cm]{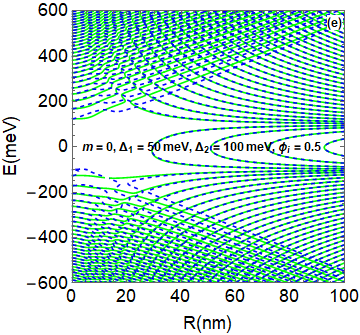}
	\includegraphics[width=5.5cm]{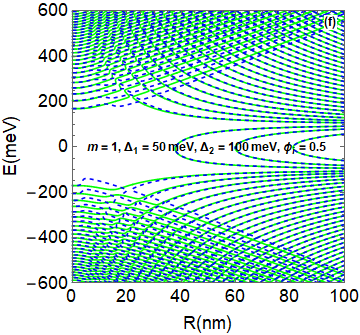}\\
	\includegraphics[width=5.5cm]{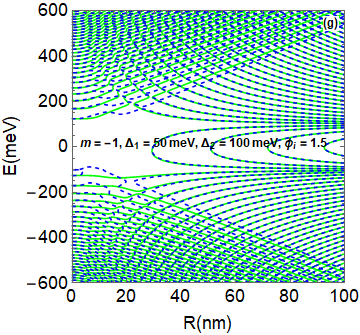}
	\includegraphics[width=5.5cm]{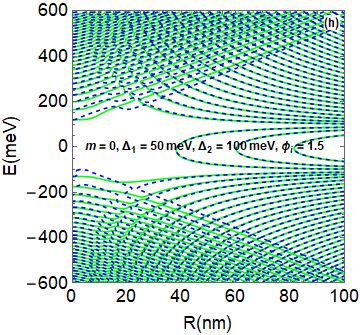}
	\includegraphics[width=5.5cm]{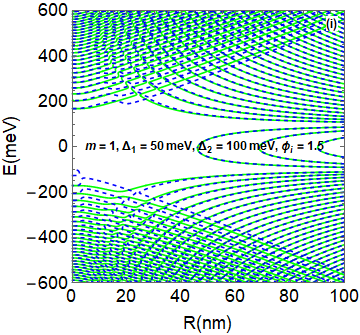}\\
	
	\caption  {(color online) The energy levels  as a function of the radius $R$
		with the same conditions as in Fig. \ref{f2}. Here green curves for  $\tau=-1$ and dashed red curves for $\tau=1$.
		 }	\label{f4}
\end{figure*}

Fig. \ref{f4} shows the energy spectrum versus radius $R$ for three AB flux values (\(\phi_i=0,0.5,1.5\)) and quantum numbers \(m\). When \(R\) is very small, we observe that the energy levels of the two valleys \(K\) and \(K'\) are not degenerate, especially when \(R \rightarrow 0\), i.e., the GMQDs disappear. As \(R\) is increased, it becomes clear that the energy levels have a linear shape, demonstrating the symmetry \(E(m, \tau) = E(m,-\tau)\), similar to that obtained in \cite{Belouad, Mirza}. The band gap also decreases, suggesting a trend where larger GMQDs tend to have a smaller disparity between the occupied electron states and the corresponding hole states \cite{Belouad, Mirza}. 
Furthermore, especially for higher values of \(R\), all figures show numerous additional levels between the two bands.
In Figs. \ref{f4}(g,h,i), an increase in the number of energy levels is shown in the region bounded by the dual gaps \(-\Delta_2\) and \(\Delta_2\), in contrast to the results reported in \cite{Belouad, Mirza}. We note that the introduction of $\phi_i$ reduces the number of energy levels present in the band gap. This decrease  becomes more significant as $\phi_i$  increases.

\begin{figure*}[ht] 
	\centering
	\includegraphics[width=5.5cm]{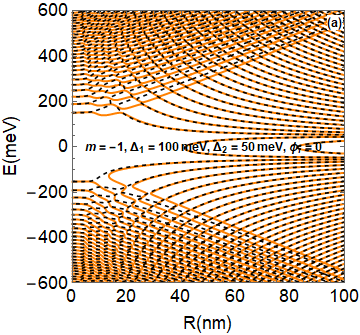}	
	\includegraphics[width=5.5cm]{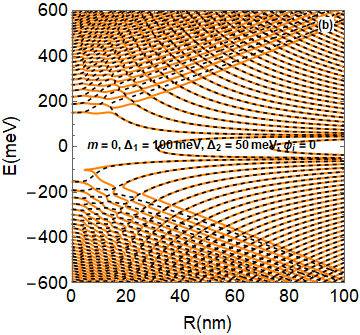}
	\includegraphics[width=5.5 cm]{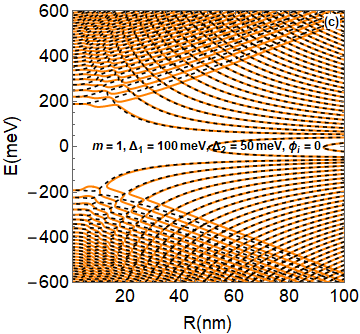}\\
	\includegraphics[width=5.5cm]{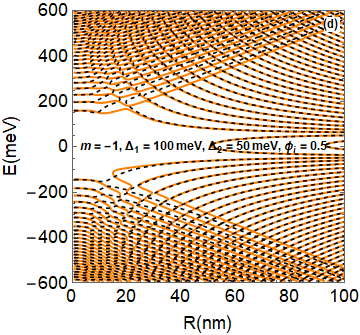}
	\includegraphics[width=5.5cm]{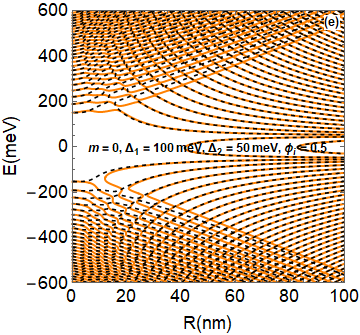}
	\includegraphics[width=5.5cm]{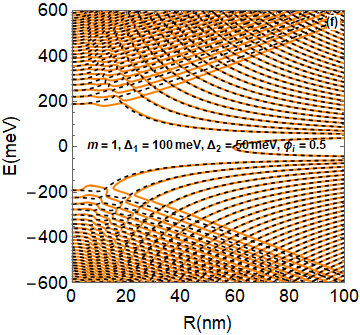}\\
	\includegraphics[width=5.5cm]{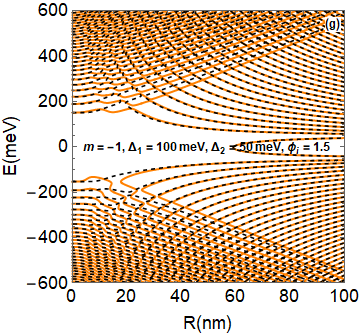}
	\includegraphics[width=5.5cm]{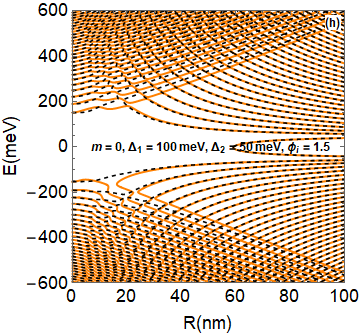}	\includegraphics[width=5.5cm]{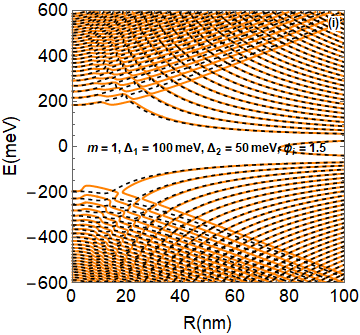}\\
	
	\caption {(color online) The energy levels  as a function of the radius $R$
		with the same conditions as in Fig. \ref{f3}. Here orange curves for $\tau=-1$ and  dashed black curves for $\tau=1$.}
	\label{f5}
\end{figure*}

 In Fig. \ref{f5}, we analyze the same scenario as in Fig. \ref{f4} by inverting the dual gaps, $\Delta_1 = 100$ meV and $\Delta_2 = 50$ meV. It is noticeable that the space between the two bands decreases, along with the levels they encompass, as the inner gap $\Delta_2$ decreases. These results are consistent with those reported in \cite{Belokda}. As the AB flux increases to $\phi_i = 0.5$ in Fig. \ref{f5}(d,e,f) and $\phi_i = 1.5$ in Fig. \ref{f5}(g,h,i), a reduction in the number of energy levels generated within the band gap is observed. There is also a noticeable decrease in the width of the band gap as $\phi_i$ increases, as shown in Fig. \ref{f5}(d,e,f). This decrease is consistent for all states of the quantum number $m$. Our analysis leads to the conclusion that the AB flux influences both the number of energy levels present and the size of the band gap. 

\begin{figure*}[ht]
	\centering
	\includegraphics[width=5.5cm]{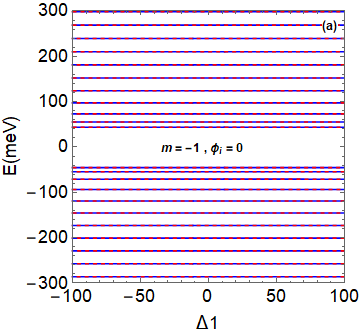}	
	\includegraphics[width=5.5cm]{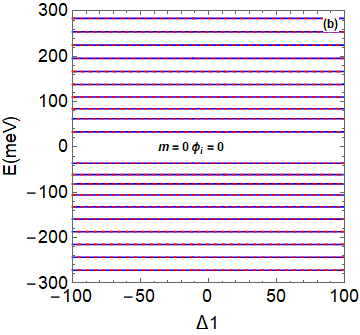}
	\includegraphics[width=5.5 cm]{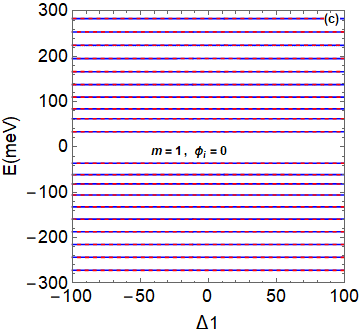}\\
	\includegraphics[width=5.5cm]{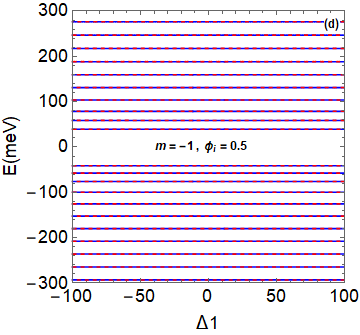}
	\includegraphics[width=5.5cm]{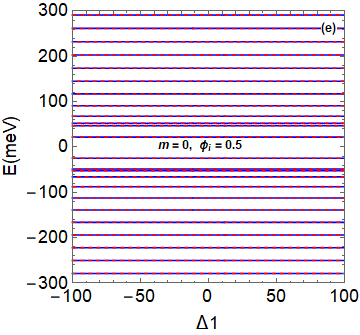}
	\includegraphics[width=5.5cm]{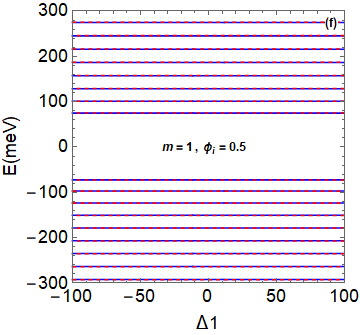}\\ 
	\includegraphics[width=5.5cm]{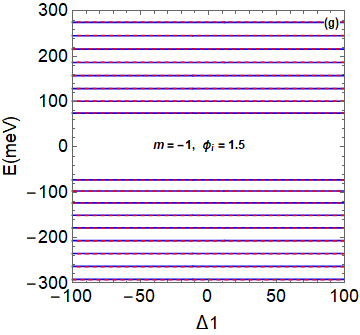}
	\includegraphics[width=5.5cm]{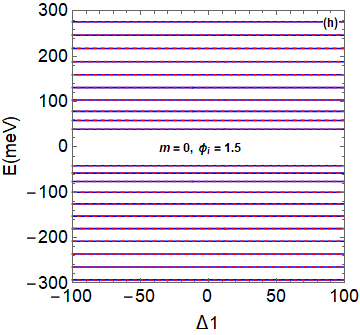}
	\includegraphics[width=5.5cm]{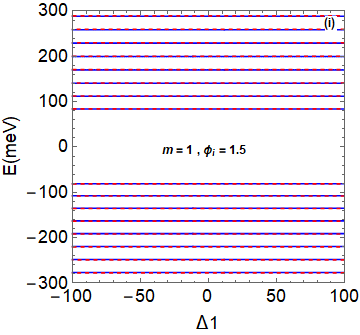}\\ 
 	\caption {(color online) The energy levels  as a function of the internal gap $\Delta_1$ for $\Delta_2=50$ meV and with the same conditions as in Fig. \ref{f2}. Here blue curves for $\tau=-1$ and dashed red curves for $\tau=1$.}
 		\label{f6}
\end{figure*}

In Fig. \ref{f6}, the energy levels are plotted against the  gap $\Delta_{1}$ within the GMQDs for three values of the quantum number $m = 0,\pm1$, with the gap $\Delta_{2}=50$ meV. The AB flux is varied as $\phi_i = 0$ for the first row, $\phi_i = 0.5$ for the second row, and $\phi_i = 1.5$ for the third row. 
In Figs. \ref{f6}(a,b,c) it is noticeable that the energy levels show a linear variation as long as $\Delta_{1}$ increases, the symmetry $E(m, \tau) = E(m, -\tau)$ is still valid. 
A distinct band gap is observed between the valence and conduction bands, the magnitude of which increases with the AB flux $\phi_i$. This trend is particularly pronounced in Figs. \ref{f6}(d,e,f) and Figs. \ref{f6}(g,h,i), highlighting how the AB flux increases the band gap except when $m = 0$ and $\phi_i = 0.5$. Such an observation indicates the influence of the AB flux on the electronic properties of the GMQDs.

\begin{figure*}[ht].    
	\centering
	\includegraphics[width=5.5cm]{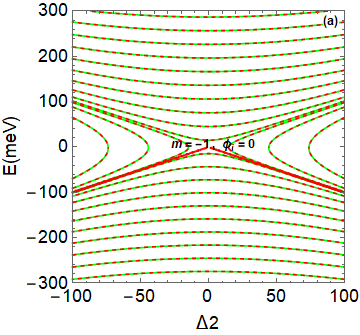}	
	\includegraphics[width=5.5cm]{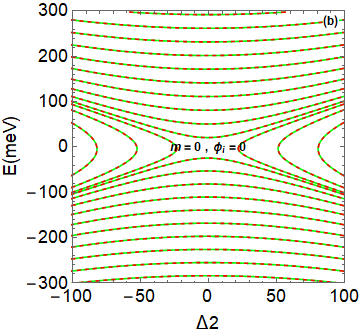}
	\includegraphics[width=5.5 cm]{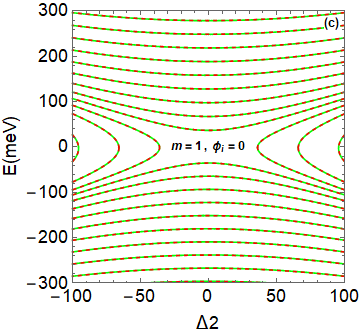}\\
	\includegraphics[width=5.5cm]{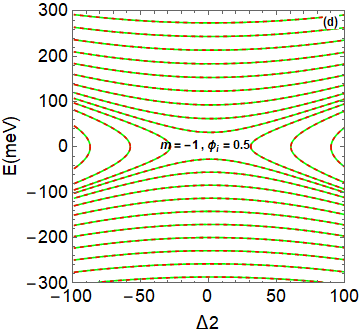}
	\includegraphics[width=5.5cm]{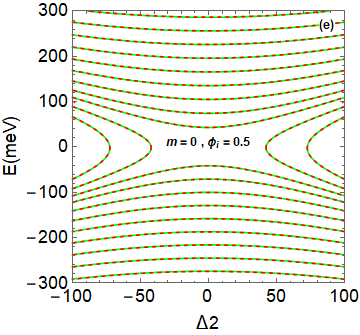}
	\includegraphics[width=5.5cm]{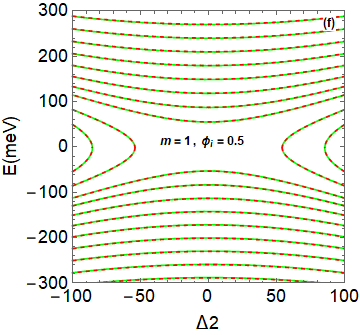}\\
	\includegraphics[width=5.5cm]{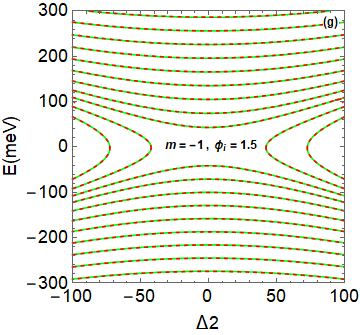}
	\includegraphics[width=5.5cm]{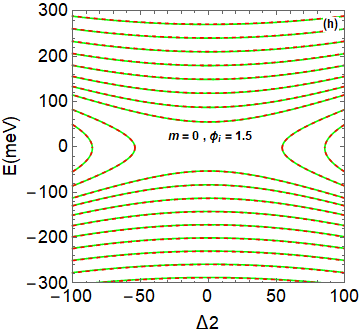}	\includegraphics[width=5.5cm]{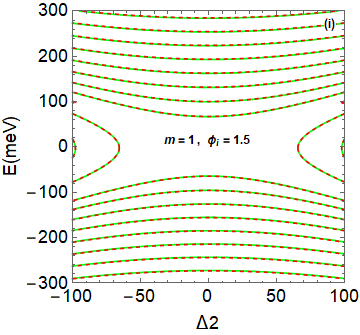}\\ 
	\caption {{The energy levels  as a function of the external gap $\Delta_2$ for $\Delta_1=50$ meV and with the same conditions as in Fig. \ref{f2}. Here green curves for $\tau=-1$ and dashed red curves for $\tau=1$.}}
	\label{f7}		
\end{figure*}

 Fig. \ref{f7} presents a comprehensive analysis of the energy levels versus the gap $\Delta_2$ outside the GMQDs, with $\Delta_1 = 50$ meV and $\phi_i=0, 0.5, 1.5$. Figs. \ref{f7}(a,b,c) clearly show that the energy levels consist of levels with a horizontal parabolic shape, reaching a minimum at $\Delta_2 = 0$. We note the appearance of additional vertical parabolic lines in the range $-2\Delta_1$ to $2\Delta_1$ ($-2\Delta_1 \leq E \leq 2\Delta_1$). These show symmetrical parallels centered around the energy point $E(m, \tau) = E(m, -\tau)$. Notably, the study reveals the emergence of additional levels within the two existing bands, confirming previous findings in research articles \cite{Belokda, Bouhlal, Azar24}. Furthermore, under the influence of the AB flux $\phi_i$, an expansion phenomenon occurs between the bands, leading to exciting new observations. Fig. \ref{f7}(d,e,f) and Fig. \ref{f7}(g,h,i) visually illustrate the effect of varying the flux on the energy levels. As $\phi_i$ increases, the number of energy levels between the bands decreases significantly, leading to a noticeable increase in the band gap, especially for certain quantum numbers $m = 0, \pm 1$. It is therefore clear that the AB flux plays a key role in controlling the band gap. Consequently, it affects the distribution of accessible energy states in the conduction and valence bands, contrary to the results presented in the previous study \cite{Belokda}.

\begin{figure*}[ht]
	\centering 
	\includegraphics[width=5.5cm]{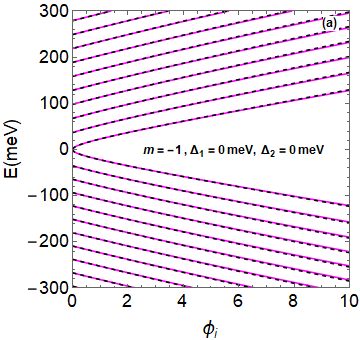}	
	\includegraphics[width=5.5cm]{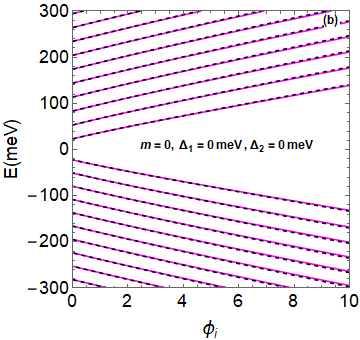}
	\includegraphics[width=5.5 cm]{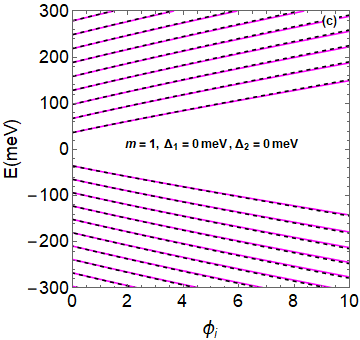}\\
	\includegraphics[width=5.5cm]{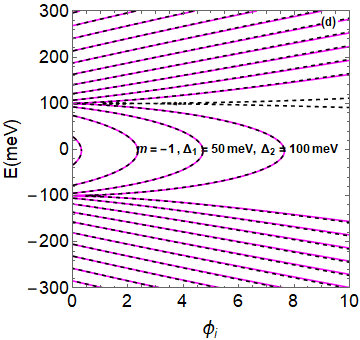}
	\includegraphics[width=5.5cm]{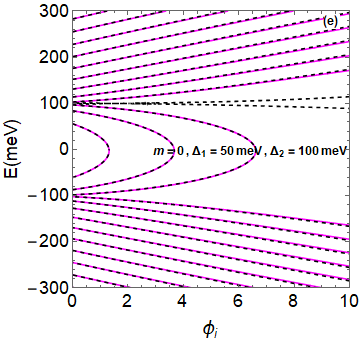} 
	\includegraphics[width=5.5cm]{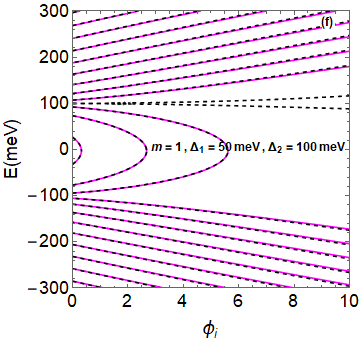}\\
	\includegraphics[width=5.5cm]{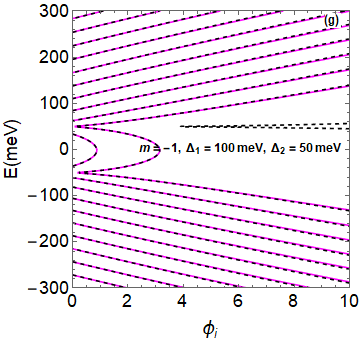}
	\includegraphics[width=5.5cm]{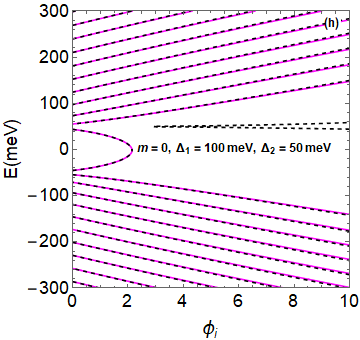}	
	\includegraphics[width=5.5cm]{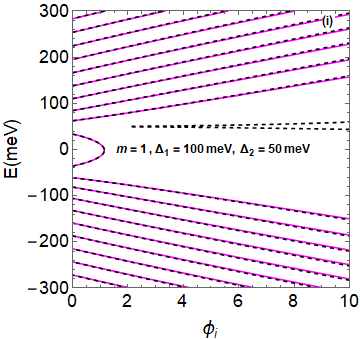}\\ 
	\caption {(color online) The energy levels as a function of the AB-flux $\phi_i$ for $R = 70$ nm , B=$10$ T and different values of $m$ and dual gaps:  (a,b,c) for ($\Delta_1=0$ meV, $\Delta_2=0$ meV), (d,e,f) for ($\Delta_1=50$ meV, $\Delta_2=100$ meV), (g,h,i) for ($\Delta_1=100$ meV, $\Delta_2=50$ meV). Different panels are also represented by changing the value of $m$, such as (a,d,g) for $m = -1$, (b,e,h) for $m = 0$ and (c,f,i) for $m = 1$. The magenta curves represent $\tau=-1$ and the dashed black curves represent $\tau=1$.}\label{f8}	
\end{figure*}

Fig. \ref{f8} shows the energy levels versus the AB flux $\phi_i$ for three quantum numbers $m=-1,0,1$. For the case of zero dual gaps in Figs. \ref{f8}(a,b,c), we observe that the energy levels exhibit a linear dependence on $\phi_i$. In addition, these energy levels exhibit the symmetry $E(m, \tau) = E(m, -\tau)$. There is also a significant band gap between the valence and conduction bands. For $\Delta_1 = 50$ meV and $\Delta_2 = 100$ meV in Figs. \ref{f8}(d,e,f), parabolic levels with a vertical shape appear within the band gap for different quantum number states. This effect is particularly pronounced for AB flux values $\phi_i \leq 6$. Interestingly, at the critical energy value $E = \Delta_1$, the energy levels are only present for $\tau = 1$, leading to a symmetry breaking. 
When the dual gaps are reversed to $\Delta_1 = 100$ meV and $\Delta_2 = 50$ meV in Figs. \ref{f8}(g,h,i), the energy levels show a similar pattern to the  two previous scenarios. 
However, there is a noticeable reduction in the number of parabolic levels within the band gap. By comparing these results with those previously reported in \cite{Bouhlal}, it can be concluded that the AB flux can be effectively used as a tool to tune both the band gap and the number of energy levels in the GMQDs.

\section{Conclusion}

\par We have studied the behavior of Dirac fermions in a graphene magnetic quantum dots (GMQDs) subjected to both an Aharonov-Bohm (AB) flux $\Phi_{AB}$ and dual gaps ($\Delta_1$ inside and $\Delta_{2}$ outside the GMQDs). In particular, we have studied how $\Phi_{AB}$ affects the energy levels of the GMQDs. To achieve this goal, we first solved the Dirac equation to determine the eigenspinors within two different regions bounded by the radius $R$ of the GMQDs. By ensuring the continuity of the eigenspinors at the boundary, we were able to derive an equation that describes the relationship between the energy levels and the physical parameters that define the GMQDs.

\par Recognizing the complexity involved in obtaining an explicit solution for the energy spectrum, we resorted to a numerical approach to elucidate the fundamental features of the system. This allowed us to study the correlation between the energy levels and several parameters, including the magnetic field $B$, the angular momentum $m$, the radius $R$, the AB flux $\phi_i$, and dual gaps ($\Delta_1$, $\Delta_2$). More precisely, we have shown how $\phi_i$, $\Delta_1$, and $\Delta_2$ can affect the energy levels of the GMQDs. Indeed, under weak $B$ we observed a degeneracy in the energy levels along with the appearance of parabolic levels within the band gap. As $B$ increased, new energy levels associated with Landau levels, typical of systems exposed to a magnetic field, began to appear. As a result, we observed that the AB flux increases the separation between the valence and conduction bands, while the energy gap breaks the symmetry of the system. 

In summary, the AB flux and the dual gaps are proving to be powerful tools for modifying the electronic properties of graphene. This controllability opens promising avenues for the design of optoelectronic, spintronic, and other quantum devices. Our study underscores the critical importance of these parameters in manipulating the electronic properties of graphene, paving the way for further fundamental and applied research in this field.


\begin{thebibliography}{99}
	

	\bibitem{Novs}K. S. Novoslov, A. K. Geim, S. V. Morozov, D. Jiang, Y. Zhang, S. V. Dubonos, I. V.
	Grigorieva, and A. A. Firsov, Science 306, 666 (2004).
	\bibitem{geim2007rise} A. K. Geim and K. S. Novoselov, 
	Nat. Mater. 6, 183 (2007).
	\bibitem{xia2010index} X.-R. Xia, N. A. Monteiro-Riviere, and J. E. Riviere, 
	Nat. Nanotechnol. 5, 671 (2010).
	
		\bibitem{guinea} A. C. Neto, F. Guinea, N. M. Peres, K. S. Novoselov, and A. K. Geim, Rev. Mod. Phys. 81, 109 (2009).
	
	 \bibitem{kats} M. Katsnelson and K. Novoselov, Solid State Communi. 143, 3 (2007).
	\bibitem{xia} X.R. Xia, N. A. Monteiro-Riviere, and J. E. Riviere, Nat. Nanotechnol. 5, 671 (2010).
	
	 \bibitem{hwang2008acoustic} E. Hwang and S. D. Sarma, 
	 Phys. Rev. B 77, 115449 (2008).
	 \bibitem{neto} J. A. Neto, M. Bueno, and C. Furtado, Ann. Phys. 373, 273 (2016). 
	 \bibitem{sprinkle} M. Sprinkle, D. Siegel, Y. Hu, J. Hicks, A. Tejeda, A. Taleb-Ibrahimi, P. Le Fevre, F. Bertran, S. Vizzini, H. Enriquez, et
	 al., Phys. Rev. Lett. 103, 226803 (2009).
	 \bibitem{zhan} D. Zhan, J. Yan, L. Lai, Z. Ni, L. Liu, and Z. Shen, Adv. Mater. 24, 4055 (2012).
	 \bibitem{rozh} A. V. Rozhkov, G. Giavaras, Yury P. Bliokha, Valentin Freilikher, and Franco Nori, Phys. Rep. 503, 77 (2011).
	 
	
	 	\bibitem{martin} A. De Martino, L. Dell’Anna, and R. Egger, Phys. Rev. Lett. 98, 066802 (2007).
	 \bibitem{espino} T. Espinosa-Ortega, I. A. Luk’yanchuk, and Y. G. Rubo, Phys. Rev. B 87, 205434 (2013).
	 \bibitem{chen} H. Y. Chen, V. Apalkov, and T. Chakraborty, Phys. Rev. Lett. 98, 186803 (2007).
	 
	 	\bibitem{giav} G. Giavaras and Franco Nori, Appl. Phys. Lett. 97, 243106 (2010); ibed, Phys. Rev. B 83, 165427 (2011); ibed,
	 Phys. Rev. B 85, 165446 (2012).
	 
	 \bibitem{zebro} D. P. Zebrowski, E. Wach, and B. Szafran, Phys. Rev. B 88, 165405 (2013).
	 \bibitem{thom} M. R. Thomsen and T. G. Pedersen, Phys. Rev. B 95, 235427 (2017).
	 \bibitem{recher} P. Recher, J. Nilsson, G. Burkard, and B. Trauzettel, Phys. Rev. B 79, 085407 (2009).
	 \bibitem{pablo} Pablo A. Denis, Chem. Phys. Lett. 492, 251 (2010).					
	 \bibitem{saeed} Saeed Sajjadi, Alireza Khataeea, Reza Darvishi Cheshmeh Soltanic, and Aliyeh
	 Hasanzadeh, J. Phys. Chem. Solids 127, 140 (2019).
	 \bibitem{kumar} Lalit Kumar Sharma, Manoranjan Kar, Ravi Kant Choubey, and Samrat Mukherjee, Chem. Phys. Lett. 780, 138902 (2021).
	 
	 	\bibitem{berger} Claire Berger, Zhimin Song, Xuebin Li, Xiaosong Wu, Nate Brown, Cécile Naud,
	 Didier Mayou, Tianbo Li, Joanna Hass, Alexei N. Marchenkov, Edward H. Conrad,
	 Phillip N. First, Walt A. de Heer, Science 312, 1191 (2006).
	 
	 \bibitem{espi} T. Espinosa-Ortega, I. A. Luk’yanchuk, and Y. G. Rubo, Phys. Rev. B 87, 205434 (2013).
	 
	 
	 \bibitem{bacon} M. Bacon, S.J. Bradley, and T. Nann, Part. Syst. Charact. 31, 415 (2013).
	 \bibitem{trauz} B. Trauzettel, D.V. Bulaev, D. Loss, and G. Burkard, Nat. Phys. 3, 192 (2007).
	 \bibitem{sun} H. Sun, L. Wu, W. Wei, and X. Qu, Mater. Today 16, 433 (2013).
	 
	 \bibitem{Aharonov1959}
	 Y. Aharonov and D. Bohm, 
	 Phys. Rev B 115, 485 (1959).
	 
	 	\bibitem{Myoung19} N. Myoung, J. W. Ryu, H. C. Park, S. J. Lee, and S.
	 Woo, Phys. Rev. B 100, 045427 (2019).
	 
	 	\bibitem{Belokda} F. Belokda, A. Jellal, and E. H. Atmani, Physica B 664, 415022 (2023).
	 	
	\bibitem{Chung2021}S. Chung, R. A. Revia, and M. Zhang,
	Adv Mater. 33, e1904362 (2021).
	
		\bibitem{ikhdair2015nonrelativistic} S. M. Ikhdair, B. J. Falaye, and M. Hamzavi, Ann. Phys. 353, 282 (2015).
		
	\bibitem{Abra} M. Abramowitz, I.A. Stegun, Handbook of Mathematical Functions (Dover Publications, Inc, New York, 1965).	
		
		
		
		
	



	

	
	

	\bibitem{Farsi21} A. Farsi, A. Belouad, and A. Jellal, Eur. Phys. J. B 94, 9 (2021).
		\bibitem{Mirza} M. Mirzakhani, M. Zarenia, S. Ketabi, D. Da Costa, and
	F. Peeters, Phys. Rev. B 93, 165410 (2016).
	

		
	
	
	\bibitem{Belouad} A. Belouad, B. Lemaalem, A. Jellal, and H. Bahlouli, Mater. Res. Express 7, 015090 (2020).
	
	\bibitem{Bouhlal} A. Bouhlal, M. El Azar, A. Siari, and A. Jellal, arXiv:2312.12324 (2023), to appear in Comput. Mater. Sci. (2024). 
	\bibitem{Azar24} M. El Azar, A. Bouhlal, and A. Jellal, Physica B 685, 416005 (2024).
		\end{thebibliography}
		 \end{document}